\documentclass[authoryear,final,3p,12pt]{elsarticle}

\usepackage{bbm}
\usepackage{url}
\usepackage{tabularx}
\usepackage{colortbl,booktabs}
\usepackage{multirow}
\usepackage{threeparttable}
\usepackage{makecell}
\usepackage{algorithm}
\usepackage{verbatim}
\usepackage{algorithmic}
\usepackage{xcolor}
\usepackage{diagbox}
\usepackage{enumerate}
\usepackage{threeparttable}
\usepackage{graphicx}
\usepackage{hyperref}
\usepackage{subfigure}

\usepackage{amsthm,amsmath,amssymb}
\newcommand{\revise}[1]{\textcolor{black}{#1}}

\journal{Elsevier}

\begin{document}

\begin{frontmatter}

\title{Probabilistic wind power forecasting resilient to missing values: an adaptive quantile regression approach}


\author[inst1]{Honglin Wen}

\affiliation[inst1]{organization={Department of Electrical Engineering, Shanghai Jiao Tong University}}

\begin{abstract}
Probabilistic wind power forecasting approaches have significantly advanced in recent decades. However, forecasters often assume data completeness and overlook the challenge of missing values resulting from sensor failures, network congestion, etc. Traditionally, this issue is addressed during the data preprocessing procedure using methods such as deletion and imputation. Nevertheless, these ad-hoc methods pose challenges to probabilistic wind power forecasting at both parameter estimation and operational forecasting stages. In this paper, we propose a resilient probabilistic forecasting approach that smoothly adapts to missingness patterns without requiring preprocessing or retraining.
Specifically, we design an adaptive quantile regression model with parameters capable of adapting to missing patterns, comprising two modules. The first is a feature extraction module where weights are kept static and biases are designed as a function of missingness patterns. The second is a non-crossing quantile neural network module, ensuring monotonicity of quantiles, with higher quantiles derived by adding non-negative amounts to lower quantiles. The proposed approach is applicable to cases under all missingness mechanisms including missing-not-at-random cases.
Case studies demonstrate that our proposed approach achieves state-of-the-art results in terms of the continuous ranked probability score, with acceptable computational cost.

\end{abstract}


\end{frontmatter}


\section{Introduction}

\subsection{Background and Motivation}
As a cornerstone of the carbon-neutral initiatives, wind power installations have surged in recent decades. However, the uncertain nature of wind power poses challenges to power system operations and electricity markets \citep{morales2013integrating}. For that, probabilistic wind power forecasting is deemed as the workhorse to accommodate the inherent uncertainty. It leverages information up to the current time and communicates the probability distribution of wind power generation at a future time in the forms of quantiles, prediction intervals, densities, etc \citep{gneiting2014probabilistic}. It has been adopted in several power system applications, including energy trading \citep{pinson2007trading} as well as reserve management \citep{matos2010setting}, and has attracted increasing interest from industries \citep{haupt2019use}.

Although forecasting approaches and products have advanced significantly through cutting-edge techniques (see a recent review by \citet{sweeney2020future}), there is often an assumption that data is complete and available during the model training and forecasting stages. However, the quality of forecasts is heavily reliant on data quality and availability. For instance, the quality of forecasts can be enhanced by incorporating nearby relevant data via various data sharing mechanisms  \citep{cavalcante2017lasso,goncalves2021privacy,goncalves2020towards}. Nevertheless, data availability is not 100\% guaranteed in industrial applications, presenting a prominent data management challenge related to missing values \citep{bohlke2020resilient}. In fact, missing values are inevitable in the energy sector, including photovoltaic and wind farm data \citep{tawn2020missing,livera2021data}, probably due to sensor failures and communication errors. Especially for offshore wind farms operating under severe adverse weather conditions, network congestion or sensor failures can often occur, leading to varying degrees of data unavailability \citep{sun2021imputation}. Even numerical weather predictions (NWPs) published by weather forecasting agencies can occasionally be a concern due to delays in provision \citep{stratigakos2023towards}.

Intuitively, the presence of missing values poses challenges at both the model training and forecasting stages. During the model training stage, missing values are commonly addressed through preprocessing, typically involving deletion or imputation \citep{liu2021two,liao2021data}. The deletion method often involves discarding samples if any part of them is missing, leading to the loss of information contained in the observed portions of deleted samples. Thus, it is only feasible when some periods of data are missing, while infeasible if data are missing sporadically. In contrast, the imputation method retains all samples, substituting missing values with estimates. While at the forecasting stage, practitioners may resort to empirical solutions, such as using the last observation or historical power distribution as forecasts when faced with missing values. Alternatively, they may need to impute the features once again to accommodate the developed model. While it is acceptable in point forecasting when missing rates are low, the quality of probabilistic forecasts are considerably degraded \citep{stratigakos2023towards,wen2023wind}. Besides, as advocated by \citet{bohlke2020resilient}, forecasting models for industrial applications should be resilient, avoiding the need for excessive manual tuning or ad-hoc solutions in the presence of missing values. Consequently, it remains an open challenge in developing probabilistic wind power forecasting approaches that are resilient to missing values.

\subsection{Related Works}
Usually, probabilistic wind power forecasting models are developed in a data-driven manner, employing either parametric (for instance \citep{zhang2020improved,wen2022sparse,dong2023transferable}) or non-parametric approaches (for instance \citep{wan2013optimal,wang2020probabilistic,yuan2021multi,dong2022regional,fan2023fluctuation}). The parametric approach assumes that wind power generation follows a specific distribution and estimates shape parameters using machine learning methods, whereas the non-parametric approach is distribution-free. Among non-parametric methods, quantile regression (QR) \citep{koenker2001quantile} is widely favored for its ease of use and state-of-the-art performance in forecasting competitions. It has been proposed to combine QR and advanced techniques like gradient boosting machine \citep{landry2016probabilistic} and online learning \citep{gallego2016line}. 
However, this approach necessitates estimating a separate model for each quantile level, often resulting in quantile-crossing phenomena where higher quantiles are smaller than lower quantiles. For that, \citet{lu2022probabilistic} have proposed a non-crossing sparse-group Lasso-quantile regressive deep neural network, incorporating constraints on the monotonicity of quantiles to mitigate quantile crossing. \citet{wen2022continuous} have proposed a continuous and distribution-free probabilistic forecasting approach, capable of predicting the entire distribution at once by transforming the base distribution to the desired one, thereby naturally avoiding quantile-crossing phenomena.

In modern statistical theory, missingness mechanisms are categorized into missing completely at random (MCAR), missing at random (MAR), and missing not at random (MNAR) cases, based on whether missingness depends on observed or missing values \citep{little2019statistical}. For example, missingness resulting from sensor failures typically falls under the MCAR mechanism, as sensor failures are mostly irrelevant to observed or missing values \citep{morshedizadeh2017application,sareen2023imputation}. MNAR cases can arise from data availability attacks \citep{pan2018cyber,xu2023availability}. For instance, attacks may occur only at high wind speeds, meaning that missingness resulting from data attacks depends on missing values.
To address the missing values issue in forecasting, a natural idea is to impute missing values before training and forecasting \citep{tawn2020missing}, known as the ``impute, then predict'' strategy. It has been extensively leveraged in recent decades, most of which resort to leverage advanced imputation techniques to recover missing values. For instance, \citet{liu2020pv,liu2022missing} have employed a super-resolution perception model to impute missing values during both model training and forecasting stages. \citet{sareen2023robust} and \citet{xu2023spatio} have respectively used the de-noising auto-encoder model and tensor decomposition technique to impute missing values before training deep-learning based forecasting models. \citet{liu2023renewable} and \citep{chen2024windfix} have introduced the self-supervised learning framework for imputation, where they mask a part of the input time series randomly, and let the model predict masked values. After that, they fine-tune the self-supervised model via the forecasting task. And, it has been proposed to perform imputation and forecasting tasks simultaneously based on deep learning techniques \citep{cao2018brits}. Missing values are imputed recursively based on the model output. This idea is adopted in the DeepAR model \citep{salinas2020deepar} and a probabilistic photovoltaic power forecasting model \citep{li2022integrated}. 
However, it is suggested that even optimal single-imputed data can lead to biased parameter estimation and prediction, as uncertainty about the missing features is discarded \citep{josse2019consistency}. 

Alternatively, one can leverage multiple imputations \citep{dempster1977maximum,van2006fully} on the training data to develop a family of models. For instance, \citet{liu2018wind} leverage multiple imputation to recover several complete datasets, and train a Gaussian process regression model based on each recovered dataset. However, this approach introduces tractability issues, as it requires training several models independently.
\citet{stratigakos2023towards} have proposed a robust optimization approach for resilient energy forecasting, which minimizes the worst-case loss when a proportion of features are missing. However, it assumes some features are never missing and requires control over the number of missing features. \citet{wen2023wind} have proposed a ``universal imputation'' strategy assuming data are MAR/MCAR, which equally treats missing values and targets and focuses on the joint distribution of features and targets. After estimating the joint distribution, the probabilistic forecasts can be derived by marginalization with respect to features. Although it demonstrates better forecasting quality compared to the commonly used ``impute, then predict'' strategy, it relies on the fully conditional specification technique, the training time of which compromises its practical applications. 
Besides, most existing works are restricted to MAR and MCAR cases. In this work, we will design a new approach, that allows unbiased parameter estimation and is applicable to cases under all missingness mechanisms.

\subsection{Proposed Method and Contributions}

Regardless of the missingness mechanism, features within historical data can be categorized into different patterns based on the locations of missing values. Particularly, for features of dimension $d$, there exist potentially $2^d$ missingness patterns. Given a missingness pattern, one can collect the corresponding observed features and targets values, and train a sub-model based on this subclass data, which is referred to as the ``retrain'' strategy by \citet{tawn2020missing}. \citet{josse2019consistency,zaffran2023conformal} have demonstrated that for a missingness pattern, one can obtain the Bayes-optimal estimate for the forecasting function (e.g., mean and quantile functions) using the common supervised learning paradigm. However, this implies an exponential increase in required sub-models, and is therefore intractable. For the aforementioned $d$-dimensional features, it may require estimating $2^d$ sub-models in the worst case. Besides, samples within each missingness pattern are severely imbalanced, with some patterns containing inadequate samples for model training. Alternatively, as suggested by \citet{bertsimas2021prediction}, it is appealing to develop models adaptive to several missingness patterns. Then, forecasters only require to develop one model that is applicable to various missingness patterns. 

In this work, we propose an adaptive quantile regression approach by designing models with parameters adaptive to missingness patterns. As missingness indicators consist of combinations of binary variables, it is infeasible to incorporate them as a part of input features and update parameters via Newton's method, as in the commonly used adaptive models \citep{gallego2016line,sommer2021online}. Instead, we multiply a part of parameters with missingness indicators via elementwise product operator \citep{le2020neumiss}. Specifically, the designed model comprises two modules. The first module, focusing on feature extraction, takes input features with potentially missing values and generates latent features. We place this module in a special neural network framework, keeping the weights static while allowing the biases to adapt to missingness patterns. And we introduce nonlinearity via missingness indicators, rather than activation functions. Particularly, we design the module as a residual model by means of skip connections, which brings more powerful representation capability.
The second module, dedicated to non-crossing quantile regression, includes a set of nonlinear multi-layer perceptrons (MLPs). It takes latent features as inputs and produces several quantiles. To prevent quantile-crossing phenomena, we adopt a multi-task framework similar to \citet{park2022learning}, where higher quantiles are derived by adding lower quantiles to non-negative increments modeled by neural networks. 
We validate the proposed approach based on data from wind toolkit \citep{draxl2015wind}, where values are removed according to designed missingness mechanisms (including MCAR and MNAR mechanisms). Case studies demonstrate that the proposed model achieves state-of-the-art in terms of continuous ranked probability score (CRPS) and exhibits good performance in sharpness and reliability (especially in MNAR cases), with efficient computational complexity.

The main contributions of the proposed model are as follows:
\begin{itemize}
\item Different from the existing ``impute, then predict'' and ``universal imputation'' strategies, the proposed approach originates from the sub-classification and adaptive regression ideas, which extends the methodology for probabilistic forecasting in the presence of missing values.
\item The proposed approach also stands out for its generality to missingness mechanisms, as it is free of assumptions on missingness mechanisms. We demonstrate the effectiveness of the proposed model in both MCAR and MNAR cases.
\item We propose a probabilistic forecasting approach resilient to missing values by designing quantile regression models with parameters adaptive to missingness patterns. It is free of any preprocessing such as deletion and imputation, and avoids excessive manual tuning in the presence of missing values.
\item Compared to the state-of-the-art approaches that are also imputation-free \citep{stratigakos2023towards,wen2023wind}, the proposed model demonstrates the better quality of forecasts and efficient computation via thorough case studies.
\end{itemize}

The paper is organized as follows. Section 2 describes the preliminaries of probabilistic wind power forecasting, missingness mechanisms, and challenges for forecasting with missing values. Section 3 formulates the problem, whereas Section 4 presents the proposed approach. Section 5 presents the setups of case studies, benchmarks, and validation metrics. Section 6 presents the results while Section 7 makes further discussion. Section 8 concludes this paper.

\textbf{Notations}: we denote random variables as uppercase letters (such as $Y$), and their realizations as lowercase letters (such as $y$). We denote time as $t$ and use it as subscripts for random variables and realizations at time $t$, for instance, $Y_t$ and $y_t$. And we use $[\cdot]$ to represent the slices of vectors and matrices. Missing values are denoted as \texttt{NA}, and the observations blurred with missing values are denoted as $\Tilde{y}_t \in \mathbb{R}\cup \texttt{NA}$. Missingness indicators are the realizations of random variable $M_t$ and denoted as $m_t\in \{0,1 \}$, where $m_t =1$ implies $\Tilde{y}_t = \texttt{NA}$ and $m_t =0$ implies $\Tilde{y}_t =y_t$. 

\section{Preliminaries}

\subsection{Probabilistic Wind Power Forecasting}
Denote the wind power generation value at time $t$ as $y_t$, the realization of a random variable $Y_t$. Probabilistic wind power forecasting aims at communicating the probabilistic distribution of wind power generation at the future time with a lead time $k$, i.e., $Y_{t+k}$.
It often relies on a model $\mathcal{M}$ with parameters $\boldsymbol{\theta}$, and leverages the information up to the current time $t$, i.e., $\boldsymbol{x}_t$, the realization of a random variable $\boldsymbol{X}_t$. The information $\boldsymbol{x}_t$ may include weather and wind power generation at the previous time. In this work, let us assume that $\boldsymbol{x}_t$ consists of lagged wind power generation values of length $h$, i.e., $\boldsymbol{x}_t=[y_{t-h+1},y_{t-h+2},\cdots,y_t]^\top \in \mathbb{R}^h$. Denote the cumulative distribution function (c.d.f) of $Y_{t+k}$ as $F_{t+k}(y)$. Then, probabilistic wind power forecasting can be described as
\begin{equation}
    \hat{F}_{t+k}(y)=F_{t+k}(y|\boldsymbol{x}_t;\mathcal{M},\boldsymbol{\theta}).
\end{equation}
The model $\mathcal{M}$ can be set as some kind of distributional model such as logit-normal or normal distributional model. Alternatively, it can be set as a group of increasing quantiles. Denote the $\alpha$-th quantile of $Y_{t+k}$ as $q_{t+k}^{(\alpha)}$, which is defined as
\begin{equation}
    q_{t+k}^{(\alpha)}=F^{-1}_{t+k}(\alpha).
\end{equation}
Then the forecast for distribution $F_{t+k}(y)$ can be also derived as
\begin{align*}
    \{\hat{q}_{t+k}^{(\alpha_1)},\hat{q}_{t+k}^{(\alpha_2)},\cdots,\hat{q}_{t+k}^{(\alpha_p)} \}, \ \alpha_1 < \alpha_2 < \cdots < \alpha_p,
\end{align*}
where $\hat{q}_{t+k}^{(\alpha_i)}$ is the estimated $\alpha_i$-th quantile. Such quantiles can be derived via quantile regression \citep{koenker2001quantile}. Let $g_\alpha(\boldsymbol{x};\boldsymbol{\theta})$ represent a quantile function, such that
\begin{equation}
\label{eq3}
    q_{t+k}^{(\alpha)}=g_
    \alpha(\boldsymbol{x}_t;\boldsymbol{\theta}).
\end{equation}
The parameter $\boldsymbol{\theta}$ can be estimated via machine learning based on historical data $\{(\boldsymbol{x}_t,y_{t+k})|t=1,2,\cdots,n \}$.
Concretely, they are estimated by minimizing the pinball loss $\ell(y_{t+k},g_\alpha(\boldsymbol{x}_t;\boldsymbol{\theta}))$ across all samples, i.e.,
\begin{equation}
    \label{eq4}\hat{\boldsymbol{\theta}}=\arg \min_{\boldsymbol{\theta}} \frac{1}{n}\sum_{t=1}^n \ell(y_{t+k},g_\alpha(\boldsymbol{x}_t;\boldsymbol{\theta}))
\end{equation}
where $\ell(y_{t+k},g_\alpha(\boldsymbol{x}_t;\boldsymbol{\theta})) $ is defined as
\begin{equation}    
   \ell(y_{t+k},g_\alpha(\boldsymbol{x}_t;\boldsymbol{\theta}))=
        \max (\alpha(y_{t+k}-g_\alpha(\boldsymbol{x}_t;\boldsymbol{\theta})),(\alpha-1)(y_{t+k}-g_\alpha(\boldsymbol{x}_t;\boldsymbol{\theta}))).
\end{equation}

\subsection{Missingness Mechanism}

In the modern statistical theory \citep{little2019statistical}, missingness mechanisms can be classified into three categories: MCAR, MAR, and MNAR. 
Intuitively, the observations for both features and targets may contain missing values. To make a notational difference, here we denote the true values of features and targets as $\boldsymbol{x}_t$ and $y_{t+k}$ respectively, while write the observations for features and targets as $\Tilde{\boldsymbol{x}}_t$ and $\Tilde{y}_{t+k}$.  We introduce a missingness indicator vector $\boldsymbol{m}_t \in \{0,1\}^h$ for $\Tilde{\boldsymbol{x}}_t$, which is the realization of a random variable $\boldsymbol{M}_t$. $\boldsymbol{m}_t[i]=0$ indicates that $\boldsymbol{x}_{t}[i]$ is observed, thus $\Tilde{\boldsymbol{x}}_t[i]=\boldsymbol{x}_{t}[i]$. In contrast, $\boldsymbol{m}_t[i]=1$ indicates $\boldsymbol{x}_{t}[i]$ is missing, thus $\Tilde{\boldsymbol{x}}_t[i]=\texttt{NA}$. In addition, we denote the observed part of $\boldsymbol{x}_t$ as $\boldsymbol{x}_{o,t}$ while the missing part of $\boldsymbol{x}_t$ as $\boldsymbol{x}_{m,t}$, and denote the corresponding random variables as $\boldsymbol{X}_{o,t}$ as well as $\boldsymbol{X}_{m,t}$ respectively.

\textit{Example}: Consider a vector
\begin{align*}
    \boldsymbol{x}_t=[0.12,0.15,0.17,0.28,0.45]^\top,
\end{align*}
the second entry of which is missing. Then the observational vector for $\boldsymbol{x}_t$ is
\begin{align*}
\Tilde{\boldsymbol{x}}_t=[0.12,\texttt{NA},0.17,0.28,0.45]^\top,
\end{align*}    
whereas the observed and missing parts of $\boldsymbol{x}_t$ are
\begin{align*}
\begin{aligned}
&\boldsymbol{x}_{o,t}=[0.12,\cdot,0.17,0.28,0.45]^\top, \\
&\boldsymbol{x}_{m,t}=[\cdot,0.15,\cdot,\cdot,\cdot]^\top,
\end{aligned}
\end{align*}
where ``$\cdot$'' means the component is not specified.

In fact, the parametric model for the joint distribution of the data sample $\boldsymbol{x}_t$ and its mask $\boldsymbol{m}_t$ can be described as
\begin{equation}
    f_{\boldsymbol{X}_t,\boldsymbol{M}_t}(\boldsymbol{x}_t,\boldsymbol{m}_t;\xi,\psi)=f_{\boldsymbol{X}_t}(\boldsymbol{x}_t;\xi)f_{\boldsymbol{M}_t}(\boldsymbol{m}_t|\boldsymbol{x}_t;\psi),
\end{equation}
where $\xi$ and $\psi$ represent the parameters of distribution for data and mask. Following \citet{little2019statistical}, we define missingness mechanisms as follows:

\textbf{Definition 1 (Missing completely at random, MCAR):} \textit{For any $\boldsymbol{m}_t\sim \boldsymbol{M}_t$, 
\begin{align*}
    f_{\boldsymbol{M}_t}(\boldsymbol{m}_t|\boldsymbol{x}_t;\psi)=f_{\boldsymbol{M}_t}(\boldsymbol{m}_t;\psi).
\end{align*}}

\textbf{Definition 2 (Missing at random, MAR):} \textit{For any $\boldsymbol{m}_t\sim \boldsymbol{M}_t$, 
\begin{align*}
    f_{\boldsymbol{M}_t}(\boldsymbol{m}_t|\boldsymbol{x}_t;\psi)=f_{\boldsymbol{M}_t}(\boldsymbol{m}_t|\boldsymbol{x}_{o,t};\psi).
\end{align*}}

\textbf{Definition 3 (Missing not at random, MNAR):} \textit{For any $\boldsymbol{m}_t\sim \boldsymbol{M}_t$, 
\begin{align*}
    f_{\boldsymbol{M}_t}(\boldsymbol{m}_t|\boldsymbol{x}_t;\psi)=f_{\boldsymbol{M}_t}(\boldsymbol{m}_t|\boldsymbol{x}_{o,t},\boldsymbol{x}_{m,t};\psi).
\end{align*}}

MCAR means the missingness is independent of the data sample, whereas MAR implies the missingness is dependent on $\boldsymbol{x}_{o,t}$, yet independent of $\boldsymbol{x}_{m,t}$. If data is not MAR, then it is MNAR. Besides, we define the missingness patterns for observational features.

\textbf{Definition 4 (Missingness pattern):} \textit{A missingness pattern for observational features is a combination of missingness indicators.}

For observational features $\Tilde{\boldsymbol{x}}_t$ of dimension $h$, there are potentially $2^h$ missingness patterns, which are written as $\mathcal{MP}_1,\mathcal{MP}_2,\cdots,\mathcal{MP}_{2^h}$.

\subsection{Challenges for Forecasting with Missing Values}

In fact, both real data and missingness mechanisms are unavailable. Instead, we are only allowed to get access to observations $\{(\Tilde{\boldsymbol{x}}_t,\Tilde{y}_{t+k})|t=1,2,\cdots,n \}$. Then, the formula (\ref{eq4}) can be naively transformed into
\begin{equation}
    \label{eq7}\hat{\boldsymbol{\theta}}=\arg \min_{\boldsymbol{\theta}} \frac{1}{n}\sum_{t=1}^n \ell(\Tilde{y}_{t+k},g_\alpha(\Tilde{\boldsymbol{x}}_t;\boldsymbol{\theta})),
\end{equation}
which is obviously intractable, as both $\Tilde{\boldsymbol{x}}_t$ and $\Tilde{y}_{t+k}$ contain missing values. For that, the ``impute, then predict'' \citep{tawn2020missing} and ``universal imputation'' \citep{wen2023wind} strategies have been proposed.

\textbf{``Impute, then predict'' strategy:} we expect to impute missing values in $\Tilde{\boldsymbol{x}}_t$ with the most plausible values via an imputation model $f_{\text{IM}}(\cdot)$. And we denote the recovered value for $\Tilde{\boldsymbol{x}}_t$ as $\hat{\boldsymbol{x}}_t$, i.e.,
\begin{align*}
    \hat{\boldsymbol{x}}_t = f_{\text{IM}}(\Tilde{\boldsymbol{x}}_t).
\end{align*}
Particularly, as $(\Tilde{\boldsymbol{x}}_t,\Tilde{y}_{t+k})$ will contribute nothing to the model training if $\Tilde{y}_{t+k}$ is missing, we select samples $\Tilde{y}_{t+k}$ where $\Tilde{y}_{t+k} = y_{t+k}$. The set of indices for the selected samples is denoted as $\mathcal{T}$. Then the model training under ``impute, then predict'' strategy reduces to
\begin{equation}
    \label{eq8}\hat{\boldsymbol{\theta}}=\arg \min_{\boldsymbol{\theta}} \frac{1}{|\mathcal{T}|}\sum_{t\in \mathcal{T}} \ell(y_{t+k},g_\alpha(\hat{\boldsymbol{x}}_t;\boldsymbol{\theta})),
\end{equation}
where $|\mathcal{T}|$ represents the cardinality of set $\mathcal{T}$. At the forecasting stage, forecasters issue forecasts by
\begin{equation}
    \hat{q}_{t+k}^{(\alpha)}=g_\alpha(\hat{\boldsymbol{x}}_t;\hat{\boldsymbol{\theta}}).
\end{equation}
Though the strategy is intuitive and simple to implement, it is suggested that even optimal single imputed data $\hat{\boldsymbol{x}}_t$ are available, the parameter estimation and prediction would be biased, since the uncertainty about the missing features is discarded \citep{josse2019consistency}.

\textbf{``Universal imputation'' strategy:} instead of modeling the probabilistic forecasting in a conditional distribution modeling framework, we model the joint distribution of features and targets $f_{\boldsymbol{X}_t,Y_{t+k}}(\boldsymbol{x}_t,y_{t+k};\mathcal{M},\boldsymbol{\theta})$ with the assumption that data are MAR. At the training stage, we estimate the parameters by using incomplete data $\{(\Tilde{\boldsymbol{x}}_t,\Tilde{y}_{t+k})|t=1,2,\cdots,n \}$ based on multiple imputation models, i.e.,
\begin{equation}
    \label{eq10}\hat{\boldsymbol{\theta}}=\arg \min_{\boldsymbol{\theta}} \frac{1}{n}\sum_{t=1}^n \log f_{\boldsymbol{X}_t,Y_{t+k}}(\Tilde{\boldsymbol{x}}_t,\Tilde{y}_{t+k};\mathcal{M},\boldsymbol{\theta}).
\end{equation}
At the forecasting stage, $y_{t+k}$ will be missing by nature. We derive the forecasts via marginalization
\begin{equation}
    f_{Y_{t+k},\boldsymbol{X}_{m,t}|\boldsymbol{X}_{o,t}}({y_{t+k},\boldsymbol{x}_{m,t}|\boldsymbol{x}_{o,t}})=\frac{f_{Y_{t+k},\boldsymbol{X}_{m,t},\boldsymbol{X}_{o,t}}({y_{t+k},\boldsymbol{x}_{m,t},\boldsymbol{x}_{o,t}};\mathcal{M},\hat{\boldsymbol{\theta}})}{\int_{\boldsymbol{x}_{m,t}} \int_{y_{t+k}} f_{Y_{t+k},\boldsymbol{X}_{m,t},\boldsymbol{X}_{o,t}}({y_{t+k},\boldsymbol{x}_{m,t},\boldsymbol{x}_{o,t}};\mathcal{M},\hat{\boldsymbol{\theta}})dy_{t+k} d\boldsymbol{x}_{m,t}},
\end{equation}
\begin{equation}
    f_{Y_{t+k}|\boldsymbol{X}_{o,t}}({y_{t+k}|\boldsymbol{x}_{o,t}})=\int_{\boldsymbol{x}_{m,t}} f_{Y_{t+k},\boldsymbol{X}_{m,t}|\boldsymbol{X}_{o,t}}({y_{t+k},\boldsymbol{x}_{m,t}|\boldsymbol{x}_{o,t}}) d\boldsymbol{x}_{m,t}.
\end{equation}
This strategy works natively and is allowed to obtain unbiased estimate for model parameters. However, it relies on the assumption of MAR mechanism and raises computational concerns to both training and forecasting stages.

Unlike the two existing strategies, in this work, we aim at developing a forecasting approach without imputation and free of any assumptions on the missingness mechanisms.

\section{Methodology} 

\begin{figure}[!ht]
\centering
\includegraphics[width=5.5in]{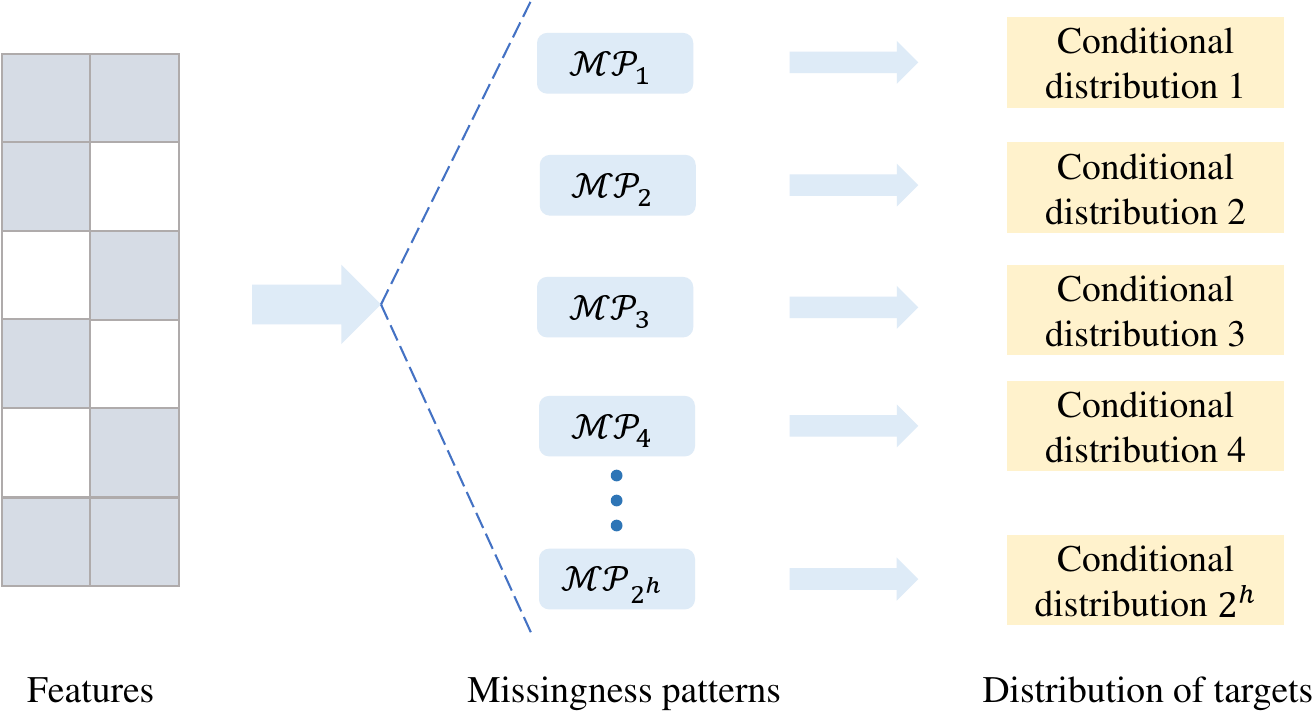}
\caption{Illustration for samples classification according to missingness patterns.}
\label{fig_partision}
\end{figure}

In this section, we illustrate our probabilistic forecasting approach directly based on observations $\{(\Tilde{\boldsymbol{x}}_t,\Tilde{y}_{t+k})|t=1,2,\cdots,n \}$. Concretely, we start with the formula (\ref{eq7}). As $(\Tilde{\boldsymbol{x}}_t,\Tilde{y}_{t+k})$ has no contributions to the parameter estimation if $\Tilde{y}_{t+k}$ is missing, we only focus on samples where $\Tilde{y}_{t+k}$ is observed in the following context. We select samples $(\Tilde{\boldsymbol{x}}_t,\Tilde{y}_{t+k})$ where $\Tilde{y}_{t+k} = y_{t+k}$ and collect the indices for the selected samples in a set, which is denoted as $\mathcal{T}$.
Then, the parameters estimation process is formally described as
\begin{equation}
    \label{eq13}\hat{\boldsymbol{\theta}}=\arg \min_{\boldsymbol{\theta}} \frac{1}{|\mathcal{T}|}\sum_{t\in \mathcal{T}}\ell(y_{t+k},g_\alpha(\Tilde{\boldsymbol{x}}_t;\boldsymbol{\theta})).
\end{equation}
Different from the formula (\ref{eq8}), here we directly deal with features $\Tilde{\boldsymbol{x}}_t$ that contain potential missing values, instead of imputed features. Following \citet{biarosenbaum1984reducing}, it is intuitive to classify $\{(\Tilde{\boldsymbol{x}}_t,y_{t+k})|t \in \mathcal{T} \}$ into several groups according to missingness patterns. Accordingly, it is allowed to estimate the distribution of the population in each group, as illustrated in Figure \ref{fig_partision}. For instance, we can collect the samples when the missingness pattern is $\mathcal{MP}_i$ as $\{(\Tilde{\boldsymbol{x}}_t,y_{t+k})|t \in \mathcal{T}, \boldsymbol{m}_t = \mathcal{MP}_i \}$. Then, it is feasible to estimate the conditional distribution 
\begin{align*}
    f_{Y_{t+k}|\boldsymbol{X}_{o,t}}({y_{t+k}|\boldsymbol{x}_{o,t};\mathcal{M}_{\mathcal{MP}_i},\boldsymbol{\theta}_{\mathcal{MP}_i},\boldsymbol{m}_t=\mathcal{MP}_i}).
\end{align*}
where the model and parameters are rewritten as $\mathcal{M}_{\mathcal{MP}_i}$ and $\boldsymbol{\theta}_{\mathcal{MP}_i}$ respectively for notational clarity.
For the considered quantile regression model (\ref{eq3}), we now formulate it as
\begin{equation}
\label{eq14}
    q_{t+k}^{(\alpha)} = g_\alpha(\boldsymbol{x}_{o,t};\boldsymbol{\theta}_{\mathcal{MP}_i},\boldsymbol{m}_t=\mathcal{MP}_i).
\end{equation}
Therefore, the quantile function can be constructed as a combination of functions.
Intuitively, it requires estimating $2^h$ sub-models for each quantile level, which scales exponentially with the dimension and is therefore impractical. 

To address the aforementioned tractability issue, we seek to develop a model whose parameters can adapt to missingness patterns and therefore write parameters as a function $\boldsymbol{\theta}(\boldsymbol{m}_t)$. Then, it is only required to establish one model for a quantile level, rather than the exponential amount of sub-models as in the formula (\ref{eq14}). 
Now the model is described as
\begin{equation}
    \label{eq15}q_{t+k}^{(\alpha)}=g_\alpha(\Tilde{\boldsymbol{x}}_t;\boldsymbol{\theta}(\boldsymbol{m}_t)).
\end{equation}
After estimating the parameters function (written as $\hat{\boldsymbol{\theta}}(\boldsymbol{m}_t)$), the model can natively leverage the observational part of $\Tilde{\boldsymbol{x}}_t$, i.e., $\boldsymbol{x}_{o,t}$ to issue forecasts.
Particularly, we design the function $g_\alpha(\Tilde{\boldsymbol{x}}_t;\boldsymbol{\theta}(\boldsymbol{m}_t))$ as a composition of two functions, namely $\phi_{\alpha,1}(\cdot)$ and $\phi_{\alpha,2}(\cdot)$, the parameters of which are respectively denoted as $\boldsymbol{\theta}_{\phi_{\alpha,1}}$ and $\boldsymbol{\theta}_{\phi_{\alpha,2}}$. And we let $\boldsymbol{\theta}_{\phi_{\alpha,1}}$ adaptive to missingness patterns while keep $\boldsymbol{\theta}_{\phi_{\alpha,2}}$ static. By denoting the output of $\phi_{\alpha,1}(\Tilde{\boldsymbol{x}}_t;\boldsymbol{\theta}_{\phi_{\alpha,1}}(\boldsymbol{m}_t))$ as $\boldsymbol{z}_t$, the model (\ref{eq15}) can be rewritten as
\begin{equation}
\label{eq16}
    \boldsymbol{z}_t =\phi_{\alpha,1}(\Tilde{\boldsymbol{x}}_t;\boldsymbol{\theta}_{\phi_{\alpha,1}}(\boldsymbol{m}_t)),
\end{equation}
\begin{equation}
    \label{eq17}q_{t+k}^{(\alpha)}=\phi_{\alpha,2}(\boldsymbol{z}_t;\boldsymbol{\theta}_{\phi_{\alpha,2}}).
\end{equation}
For computational ease, we replace \texttt{NA} in $\Tilde{\boldsymbol{x}}_t$ with 0, which operates as
\begin{align*}
    \Tilde{\boldsymbol{x}}_t \odot (1-\boldsymbol{m}_t),
\end{align*}
where $\odot$ is an elementwise product operator.

\begin{figure}[!ht]
\centering
\includegraphics[width=5.5in]{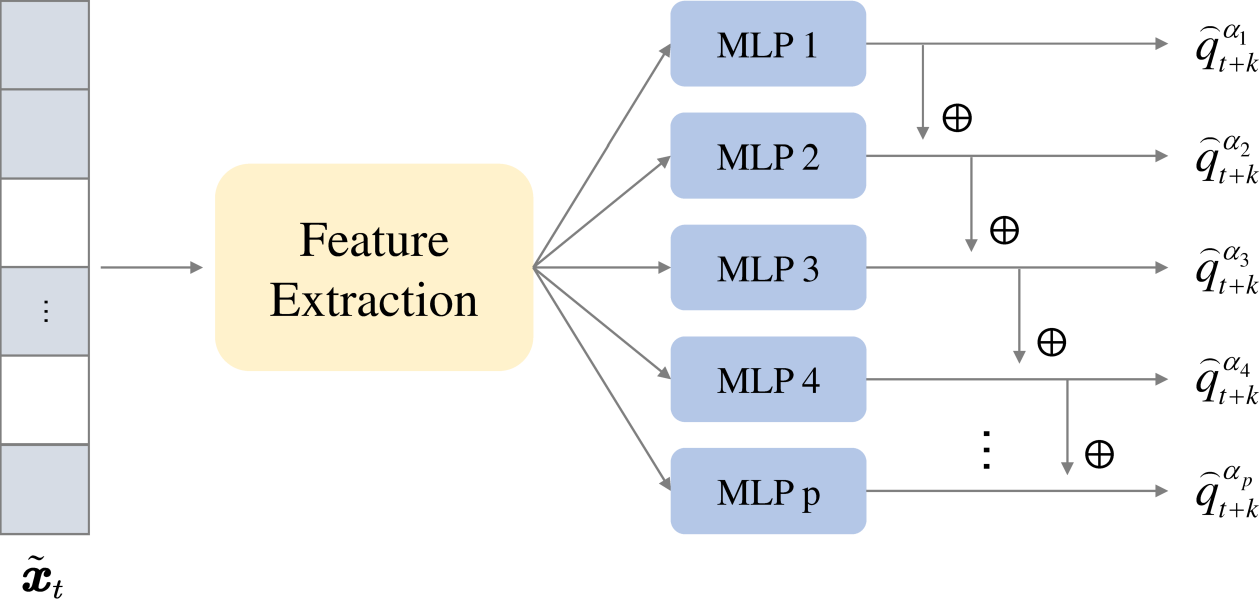}
\caption{The sketch of the proposed approach, where blank blocks in $\Tilde{\boldsymbol{x}}_t$ indicate missing values.}
\label{fig_sketch}
\end{figure}

\section{Proposed Model}

With the main idea described in Section 3, we now describe the proposed model in detail. Concretely, the function $\phi_{\alpha,1}(\cdot)$ is implemented similar to the NeuMiss block proposed by \citet{le2020neumiss}, whereas the function $\phi_{\alpha,2}(\cdot)$ is implemented as a non-crossing quantile regression neural network. In particular, we place the quantile regression model in a multi-task framework and ensure the monotonicity of quantiles. Consequently, all quantile functions share the same feature extraction block and differ in the subsequent nonlinear mappings. We sketch the proposed approach in Figure~\ref{fig_sketch}, which consists of a feature extraction block and several MLPs correlated with previous ones via addition successively.

\subsection{Feature Extraction Module}
\begin{figure}[!ht]
\centering
\includegraphics[width=2.5in]{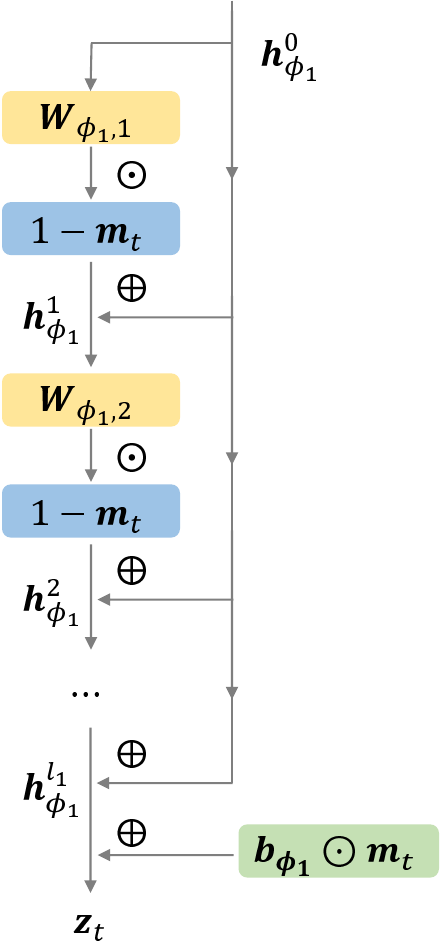}
\caption{The structure of feature extraction block.}
\label{fig_neumiss}
\end{figure}

Here we illustrate how to develop the feature extraction module via a special neural network.
Following \citet{bertsimas2021prediction}, we further specify the weights in the neural network as static whereas the biases as the functions of missingness patterns. As all quantile functions share the same feature extraction module, we drop the quantile level notation $\alpha$. Take a neural network layer as an example. The weight is denoted as $\boldsymbol{W}_{\phi_1}$ since it is kept static, while the bias is
denoted as $\boldsymbol{b}_{\phi_1}(\boldsymbol{m}_t)$. Specifically, the bias is designed as
\begin{equation}
    \boldsymbol{b}_{\phi_1}(\boldsymbol{m}_t)=\boldsymbol{b}_{\phi_1}\odot \boldsymbol{m}_t.
\end{equation}
In this work, we design a deep neural network for the feature extraction module similar to the NeuMiss model proposed by \citet{le2020neumiss}. It stacks $l_1$ blocks via skip connections, which is illustrated in Figure~\ref{fig_neumiss}. In contrast to the commonly used neural network layers, the designed blocks exclude biases and nonlinear activation functions. Alternatively, each block introduces nonlinearity by multiplying with missingness indicators $(1-\boldsymbol{m}_t)$, and contain a term from initial inputs $\boldsymbol{h}^0_{\phi_1} = \Tilde{\boldsymbol{x}}_t \odot (1-\boldsymbol{m}_t)$ via skip connections.
Then, for the $l$-th block, we denote its input as $\boldsymbol{h}^{l-1}_{\phi_1}$, and its output as $\boldsymbol{h}^l_{\phi_1}$. It operates as
\begin{equation}
    \boldsymbol{h}^l_{\phi_1} = \boldsymbol{W}_{\phi_1}^l\boldsymbol{h}^{l-1}_{\phi_1}\odot(1-\boldsymbol{m}_t)+\boldsymbol{h}^0_{\phi_1},
\end{equation}
where $\boldsymbol{W}_{\phi_1}^l$ is the weight matrix in the $l$-th block. Specially, the output of the last block $\boldsymbol{h}^{l_1}_{\phi_1}$ is derived as
\begin{equation}
    \boldsymbol{h}^{l_1}_{\phi_1} = \boldsymbol{W}_{\phi_1}^{l_1}\boldsymbol{h}^{l_1-1}_{\phi_1}+\boldsymbol{h}^0_{\phi_1},
\end{equation}
Then, the latent features $\boldsymbol{z}_t$ is derived by adding $\boldsymbol{h}^{l_1}_{\phi_1}$ with $\boldsymbol{b}_{\phi_1}\odot \boldsymbol{m}_t$, i.e.,
\begin{equation}
\label{eq24}
    \boldsymbol{z}_t = \boldsymbol{h}^{l_1}_{\phi_1}+\boldsymbol{b}_{\phi_1}\odot \boldsymbol{m}_t.
\end{equation}
It will be further fed into a group of nonlinear functions to derive quantiles as described by the formula (\ref{eq17}).

\subsection{Non-crossing Quantile Regression Neural Network}

Specifically, we define a non-negative nonlinear function $\phi_{\alpha_i,2}$ for each quantile level $\alpha_i$, with the parameters defined as $\boldsymbol{\theta}_{\phi_{\alpha_i,2}}$. Then all quantile functions are derived successively from lower quantile levels to higher quantile levels. The quantile function at the lowest quantile level, i.e., $q_{t+k}^{\alpha_1}$ is simply derived by the composition of $\phi_1(\cdot)$ and $\phi_{\alpha_1,2}(\cdot)$ as described by the formula (\ref{eq17}). It is described as
\begin{equation}
\label{eq25}
    q_{t+k}^{(\alpha_1)} =\phi_{\alpha_1,2}(\boldsymbol{z}_t;\boldsymbol{\theta}_{\phi_{\alpha_1,2}}),
\end{equation}
where $\boldsymbol{z}_t$ is derived from the formula (\ref{eq24}).
As for the quantile level $\alpha_i \ (i>1)$, we derive the quantile function $q_{t+k}^{(\alpha_i)}$ by adding  $q_{t+k}^{(\alpha_{i-1})} $ with a non-negative amount. The non-negative amount is modeled as the composition of $\phi_1(\cdot)$ and $\phi_{\alpha_i,2}(\cdot)$, i.e.,
\begin{equation}
\label{eq26}
    q_{t+k}^{(\alpha_i)} =q_{t+k}^{(\alpha_{i-1})}+\phi_{\alpha_i,2}(\boldsymbol{z}_t;\boldsymbol{\theta}_{\alpha_i,2}).
\end{equation}
As $\phi_{\alpha_i,2}$ is non-negative, we can ensure
\begin{align*}
    q_{t+k}^{(\alpha_{i-1})} \leq q_{t+k}^{(\alpha_i)}.
\end{align*}

Each function $\phi_{\alpha_i,2}$ is implemented with an $l_2$-layer MLP, where each layer takes $\boldsymbol{h}^{l-1}_{\alpha_i,2}$ as the input, and yield $\boldsymbol{h}^l_{\alpha_i,2}$ as the output. Specially, $\boldsymbol{z}_t$ derived from the formula (\ref{eq24}) is denoted as $\boldsymbol{h}^{0}_{\alpha_i,2}$ and the final output is $\phi_{\alpha_i,2}(\boldsymbol{z}_t;\boldsymbol{\theta}_{\alpha_i,2})$. The $l$-th layer operates as
\begin{equation}
    \boldsymbol{h}^l_{\alpha_i,2} = \boldsymbol{W}_{\alpha_i,2}^l \boldsymbol{h}^{l-1}_{\alpha_i,2}+\boldsymbol{b}_{\alpha_i,2}^l,
\end{equation}
where $\boldsymbol{W}_{\alpha_i,2}^l$ and $\boldsymbol{b}_{\alpha_i,2}^l$ respectively represent the weight and bias in this layer. It is followed by a Relu function $\sigma(\cdot)$ , which operates as
\begin{equation}
    \sigma(\boldsymbol{h}^l_{\alpha_i,2}) = \max (\boldsymbol{h}^l_{\alpha_i,2},\boldsymbol{0}),
\end{equation}
where $\max$ returns the maximum between $\boldsymbol{h}^l_{\alpha_i,2}$ and $\boldsymbol{0}$ elementwisely.

To estimate all the parameters, we minimize the total loss, i.e.,
\begin{equation}
 \hat{\boldsymbol{\theta}}=\arg \min_{\boldsymbol{\theta}} \frac{1}{|\mathcal{T}|}\sum_{t\in \mathcal{T}}\sum_{\alpha_i}\ell(y_{t+k},g_{\alpha_i}(\Tilde{\boldsymbol{x}}_t;\boldsymbol{\theta})),
\end{equation}
where $g_{\alpha_i}(\Tilde{\boldsymbol{x}}_t;\boldsymbol{\theta})$ is derived via formulas (\ref{eq24}), (\ref{eq25}), (\ref{eq26}), and $\boldsymbol{\theta}$ gathers all the parameters in the proposed model.

\section{Case Study}
In this work, similar to \citet{liu2018wind,liu2023renewable}, we focus on very short-term (1-3 hours ahead) wind power forecasting at a single site in the presence of missing values, where missingness often occurs due to network congestion or sensor failures \citep{sun2021imputation} as well as potential data attacks \citep{xu2023availability}. But we note that the proposed approach is general, and is allowed to be easily applied to the cases where NWPs are not provided in time \citep{stratigakos2023towards} and where data from nearby site are shared \citep{goncalves2020towards}.
In what follows, we describe the data for case validation, which come from the open-source wind toolkit \citep{draxl2015wind}. Then we 
introduce experimental setups, where missingness is simulated based on different mechanisms. After that, we describe benchmark models and qualification metrics. All experiments run on a computer with 12-core CPU and a NVIDIA GEFORCE GTX 1080 Ti GPU.

\subsection{Data Description}
As we focus on very short-term wind power forecasting, here we use a wind power measurement dataset collected from a wind farm located in South Carolina. The dataset is hourly, covering the period from 2007 to 2013, and the values are normalized by the farm's capacity. In each case, we split the first $70\%$ of data for training models, the following $10\%$ of data for validation, and the last $20\%$ of data for genuine forecast verification.

\subsection{Experimental Setups}
\begin{figure}[!ht]
\centering
\subfigure[Sporadic missingness]{\includegraphics[width=0.48\textwidth]{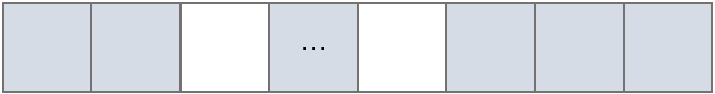}}
\subfigure[Block-wise missingness]{\includegraphics[width=0.48\textwidth]{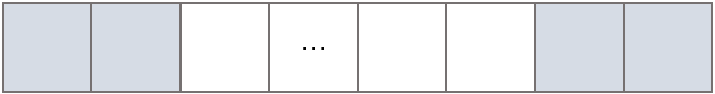}}
\caption{Illustration of two kinds of missingness cased by the MCAR mechanism, namely (a) sporadic missingness (b)block-wise missingness, where grey blocks indicate observations and white blocks indicate missingness.}
\label{fig_missingness}
\end{figure}

In this work, we verify the proposed approach on MCAR and MNAR cases, as MCAR cases are the most prevalent in the energy sector; while the MNAR case is used to demonstrate the superiority of the proposed approach compared to the state-of-the-art approaches. Then, we simulate missingness caused by MCAR and MNAR mechanisms respectively. For the MCAR mechanism, we examine sporadic and block-wise patterns of missingness, illustrated in Figure \ref{fig_missingness}. Despite the distinct patterns of missingness distribution in these two cases, it remains independent of the wind power generation value. As for the MNAR mechanism, we consider a self-masking case caused by data availability attack \citep{xu2023availability}, where values exceeding a threshold are rendered missing. 

The designed 3 cases are described as follows:

\begin{description}
    \item[Case 1:] In this case, we set the missing rate as 20\%, and randomly remove 20\% of data, which locate sporadically over the dataset.
    \item[Case 2:] In this case, we introduce missingness randomly across the dataset by removing data in contiguous blocks. The length of each block follows a uniform distribution ranging from 5 to 30 steps. Specifically, we designate the total number of blocks as 300.
    \item[Case 3:] In this case, we set the threshold of a self-masking case as 0.87, and remove data points with values exceeding this threshold.
\end{description}

In each case, we consider the lead time of 1, 2, and 3, and use wind power generation values at previous time as features. While our primary focus is not feature selection in this study, we empirically opt for 6 lags. Certainly, more sophisticated feature selection approaches are applicable.

\subsection{Benchmark Models}
In this work, we mainly consider three types of models as benchmarks: naive models, ``impute, then predict'' strategy-based models, and ``universal imputation'' strategy-based models. The naive model is set as the climatology model, which is often used by forecasters when confronted with missing values. For the ``impute, then predict'' strategy-based models, we leverage MissForest \citep{stekhoven2012missforest} to impute missing values, and develop a parametric as well as a non-parametric model based on the imputed data. Also, the DeepAR model \citep{salinas2020deepar} that uses the intermediate results of recurrent neural networks to impute missing values is considered. As for the ``universal imputation'' strategy-based model, we consider the fully conditional specification model \citep{van2006fully} used by \citet{wen2023wind}. Besides, we set a quantile regression model trained on complete data as a reference. We describe them as follows:

\begin{description}
    \item[Climatology:] It is a naive model that issues forecasts as the empirical distribution of wind power generation estimated based on all historical samples. Intuitively, it uses no contextual information at the forecasting stage, and issues the same forecasts anytime. 
    \item[IM-Gaussian:] It is an ``impute, then predict'' strategy-based model. Missing values are imputed via MissForest, based on which a parametric probabilistic forecasting model with Gaussian distributional assumption is developed. At the forecasting stage, features will missing values are imputed at first, and then fed into the forecasting model.
    \item[IM-QR:] It is an ``impute, then predict'' strategy-based model. Missing values are also imputed via MissForest, based on which a group of quantile regression models are developed. At the forecasting stage, features will missing values are imputed at first, and then fed into the forecasting model.
    \item[DeepAR:] It is a state-of-the-art ``impute, then predict'' strategy-based model, where missing values are imputed via the intermediate results of recurrent neural networks at both training and forecasting stages.
    \item[UI:] It is a ``universal imputation'' strategy-based model based on the fully conditional specification model. At the training stage, it estimates the joint distribution of features and targets based on data with missing values. Particularly, it specifies the joint distribution via the conditional distribution of each feature. At the forecasting stage, both missing features and targets are predicted via the learned distribution model. The forecasts are further derived by marginalization.
    \item[R-QR:] It is a group of quantile regression model trained on the complete dataset, which serves as a reference. It is expected models whose predictive performance is close to the reference model is preferred.
\end{description}
In addition, we consider 19 quantile levels for all quantile regression models from 5\% to 95\% with 5\% increasing pace to approximate the distribution. The hyperparameters of the benchmark and proposed model are tuned via cross validation, which are listed in appendix.

\subsection{Qualification Metrics}
To assess the quality of forecasts, we verify the calibration and sharpness of forecasts via reliability diagrams and prediction width respectively, as well as a skill score, namely continuous ranked probability score (CRPS). 
\begin{enumerate} [(1)]
    \item Calibration: The calibration of predictive densities is assessed with reliability diagrams, which presents the frequencies of conditional events against forecast probabilities. Particularly, we plot the frequencies of coverage for a quantiles from 5\% to 95\% against their nominal probabilities.
    \item Sharpness: The sharpness of forecasts is assessed with the average width of central prediction intervals, which reveals how predictive densities concentrate the information. Here we plot the width for 10\%-90\% prediction intervals.
    \item CRPS:  Given the lead time $k$, we denote the predictive c.d.f for wind power generation at time $t+k$ as $\hat{F}_{t+k}$, and the real generation value as $y_{t+k}$. The CRPS is calculated via
\begin{align*}
    {\rm CRPS}(\hat{F}_{t+k},y_{t+k})=\int_y (\hat{F}_{t+k}(y)-\mathcal{I}(y-y_{t+k}))^2 dy,
\end{align*}
where $\mathcal{I}(\cdot)$ is a step function.
We report the average CRPS of all test samples for each lead time.
\end{enumerate}
For further information on forecast verification, we refer readers to \citep{gneiting2014probabilistic}.

\section{Results}

\subsection{Case 1}
\begin{figure}[!ht]
    \centering
    \includegraphics[width=4.5in]{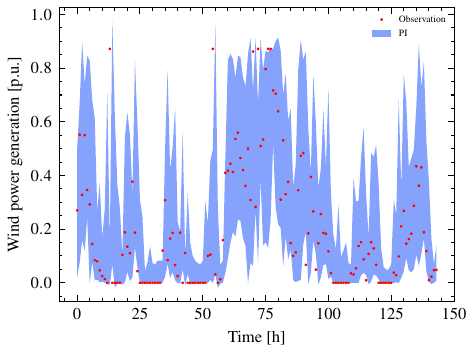}
    \caption{1-step ahead 90\% prediction intervals by the proposed model for 144 successive observations in case 1.}
    \label{fig:curve}
\end{figure}

The CRPS values for forecasts from both the proposed and benchmark models are presented in Table~\ref{table_1}. \revise{Note that the CRPS values evaluate the area between the predictive c.d.f of wind power and the observed one, and are normalized by the wind power capacity here. In reality, the generated scenarios/quantiles may differ in the values of several MW, which will have a considerable impact on energy trading and energy dispatch \citep{morales2013integrating}.} Notably, climatology exhibits the poorest performance among all models, as it relies solely on an empirical distribution without incorporating contextual information. In contrast to the common situations where quantile regression models often outperform Gaussian distributional models, The performance of IM-Gaussian and IM-QR is closely matched. Although the impact of imputation on the training of downstream forecasting models can be intricate, we infer that the accumulation of errors from several independently trained quantile regression models may be a potential cause. Surprisingly, among the three models employing the ``impute, then predict" strategy, DeepAR exhibits the poorest performance. Specifically, DeepAR imputes missing values by leveraging the intermediate results of the recurrent neural network during training, potentially leading to imputed values deviating further from real values compared to those derived from the MissForest model. The performance of UI and the proposed model is comparable to the reference model. Notably, the UI model slightly surpasses the reference model, suggesting its potential robustness against overfitting. This underscores the ancillary effects of missing values on forecasting model development, aligning with \citep{breiman2001random} which advocates subsampling feature subsets for ensemble use to enhance robustness. As an illustration, Figure~\ref{fig:curve} showcases the 1-step ahead 90\% prediction intervals generated by the proposed model for 144 consecutive observations.

\begin{figure}[!ht]
\centering
\subfigure[Reliability diagrams]{\includegraphics[width=0.48\textwidth]{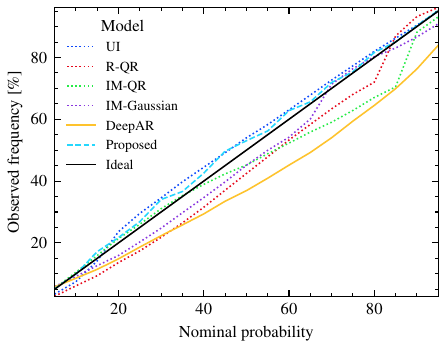}}
\subfigure[Sharpness diagrams]{\includegraphics[width=0.48\textwidth]{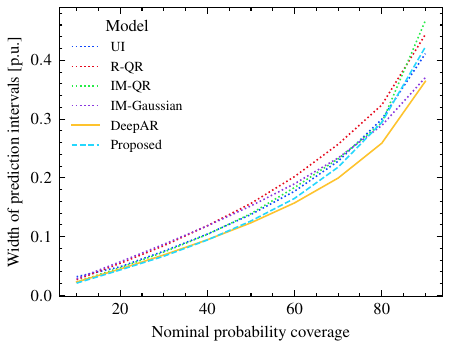}}
\caption{Assessment of 1-step ahead probabilistic forecasts for all models for Case 1, based on reliability diagrams (a) and sharpness diagrams (b).}
\label{fig_score_1}
\end{figure}

We present the reliability diagrams and prediction interval widths for 1-step ahead forecasts in Figure~\ref{fig_score_1}. As shown, the reliability diagrams of models employing the``impute, then predict" strategy deviate noticeably from the ideal case. The prediction interval widths of the proposed and benchmark models are comparable and consistently smaller than that of the reference model. DeepAR exhibits the poorest reliability among all models, although its prediction interval widths are the smallest. The reliability diagrams of UI and the proposed model closely align with the ideal case, indicating small biases in their forecasts. Surprisingly, the reliability of the reference model is inferior to that of the proposed and UI models, despite the reference model yielding larger prediction interval widths. This discrepancy may be attributed to overfitting in the reference model.

\begin{table}[!ht]
\renewcommand{\arraystretch}{1.3}
\caption{The CRPS values of forecasts by the proposed and benchmark models with different lead times in case 1 (\%).}
\label{table_1}
\centering
\begin{threeparttable}
 \begin{tabular}{cccccccc}
 \hline
 Lead time & Climatology & IM-Gaussian & IM-QR & DeepAR & UI & R-QR & Proposed\\
\hline

1 & 18.6 & 7.5 & 7.8 & 7.8 & 6.9 & 6.9 & 7.2\\
2 & 18.6 & 9.9 & 9.9 & 10.2 & 9.1 & 9.3 & 9.4\\
3 & 18.6 & 11.7 & 11.7 & 12.1 & 10.9 & 11.2 & 11.4\\

\hline
\end{tabular}
\end{threeparttable}
\end{table}

\subsection{Case 2}
Table~\ref{table_2} displays the CRPS values for forecasts generated by both the proposed and benchmark models. In this case, the differences in CRPS values among all models are smaller compared to those in case 1. Unlike case 1, missingness occurs in blocks, resulting in a greater number of samples with complete observations. Consequently, the impact of missing values on the quality of forecasts is reduced. Among models employing the ``impute, then predict" strategy, DeepAR continues to exhibit the poorest performance, although the difference between DeepAR and IM-Gaussian/IM-QR is smaller than in case 1.  In contrast, the performance of the proposed and UI models remains superior to that of ``impute, then predict" strategy-based models and is comparable to the reference model. This implies the applicability of the proposed and UI models to cases with both sporadic and block-wise missingness.

\begin{figure}[!ht]
\centering
\subfigure[Reliability diagrams]{\includegraphics[width=0.48\textwidth]{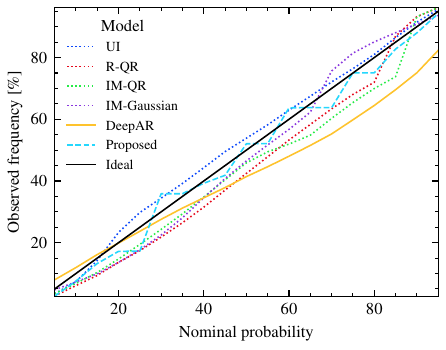}}
\subfigure[Sharpness diagrams]{\includegraphics[width=0.48\textwidth]{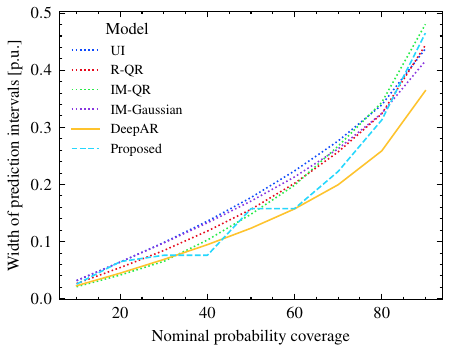}}
\caption{Assessment of 1-step ahead probabilistic forecasts for all models for Case 2, based on reliability diagrams (a) and sharpness diagrams (b).}
\label{fig_score_2}
\end{figure}

Figure~\ref{fig_score_2} illustrates the reliability diagrams and prediction interval widths for 1-step ahead forecasts. In this case, DeepAR continues to demonstrate the least reliability among all models. Notably, the reliability diagram of the proposed model fluctuates around the ideal case, albeit in close proximity. This behavior is attributed to the monotonicity constraint imposed on the proposed model. While these constraints ensure that higher quantiles are no smaller than lower quantiles, they concurrently impact parameter estimation. A more in-depth analysis is provided in the subsequent subsection. Overall, the performance of the proposed model in terms of reliability and sharpness is sound.

\begin{table}[!ht]
\renewcommand{\arraystretch}{1.3}
\caption{The CRPS values of forecasts by the proposed and benchmark models with different lead times in case 2 (\%).}
\label{table_2}
\centering
\begin{threeparttable}
\begin{tabular}{cccccccc}
\hline
Lead time & Climatology & IM-Gaussian & IM-QR & DeepAR & UI & R-QR & Proposed\\
\hline
1 & 18.6 & 7.2 & 7.4 & 7.4 & 6.6 & 6.3 & 6.4\\
2 & 18.6 & 9.6 & 9.6 & 9.9 & 8.9 & 9.2 & 9.2\\
3 & 18.6 & 11.5 & 11.5 & 11.7 & 11.9 & 11.2 & 11.2\\
\hline
\end{tabular}
\end{threeparttable}
\end{table}

\subsection{Case 3}
Table~\ref{table_3} presents the CRPS values for forecasts generated by both the proposed and benchmark models. In comparison to cases 1 and 2, all models exhibit improved CRPS performance, attributed to the exclusion of high wind power generation values that often result in substantial forecast errors. Nonetheless, the DeepAR model continues to display the poorest forecast quality among all models. The performance of UI and the proposed models surpasses that of models following the ``impute, then predict" strategy. Notably, the proposed model attains the highest performance among all models, indicating its applicability for MNAR cases.

\begin{figure}[!ht]
\centering
\subfigure[Reliability diagrams]{\includegraphics[width=0.48\textwidth]{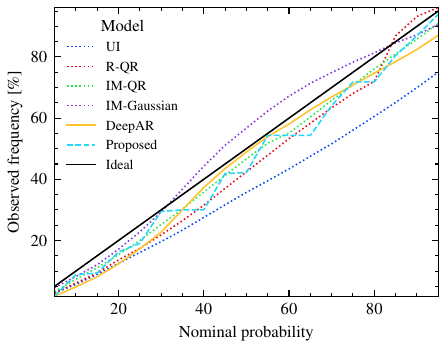}}
\subfigure[Sharpness diagrams]{\includegraphics[width=0.48\textwidth]{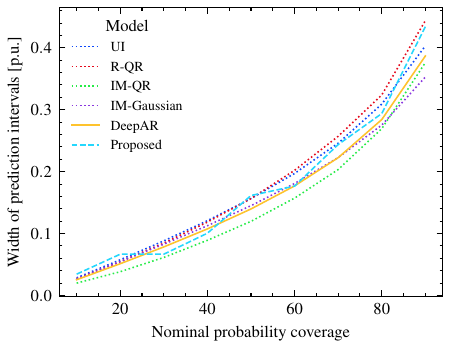}}
\caption{Assessment of 1-step ahead probabilistic forecasts for all models for Case 3, based on reliability diagrams (a) and sharpness diagrams (b).}
\label{fig_score_3}
\end{figure}

Figure~\ref{fig_score_3} illustrates the reliability diagrams and prediction interval widths for 1-step ahead forecasts. In contrast to cases 1 and 2, the reliability of the UI model deviates significantly from the ideal case. This deviation may be attributed to the case's violation of the MAR/MCAR assumption upon which the UI model relies. Conversely, the proposed model demonstrates greater suitability for MNAR cases, as it operates without such assumptions.

\begin{table}[!ht]
\renewcommand{\arraystretch}{1.3}
\caption{The CRPS values of forecasts by the proposed and benchmark models with different lead times in case 3 (percentage).}
\label{table_3}
\centering
\begin{threeparttable}
\begin{tabular}{cccccccc}
\hline
Lead time & Climatology & IM-Gaussian & IM-QR & DeepAR & UI & R-QR & Proposed\\
\hline
1 & 15.4 & 6.6 & 6.8 & 7.1 & 6.4 & 6.0 & 6.0\\
2 & 15.4 & 8.6 & 8.7 & 9.3 & 8.3 & 7.9 & 8.0\\
3 & 15.4 & 10.1 & 9.9 & 10.8 & 9.5 & 9.4 & 9.4\\
\hline
\end{tabular}
\end{threeparttable}
\end{table}

\section{Discussion}
In this section, we discuss the relationship with existing approaches, the architecture of the proposed model, and model complexity.

\subsection{Discussion on the Relationship to Existing Strategies}

The results across all cases consistently demonstrate that the proposed model attains state-of-the-art performance, whereas models based on the ``impute, then predict" strategy consistently exhibit inferior performance. Notably, the ``universal imputation" strategy-based model performs well in cases 1 and 2. It is worth mentioning that the proposed model, described in the formula (\ref{eq13}), shares several conceptual similarities with the ``impute, then predict" strategy outlined in the formula (\ref{eq8}). Both are formulated within a commonly used conditional distribution modeling framework, differing primarily in their treatment of missing values. The ``impute, then predict'' strategy aims to recover missing values and then fit off-the-shelf forecasting models, whereas the proposed model draws inspiration from the retraining strategy discussed by \citet{tawn2020missing,stratigakos2023towards}, often regarded as a strong benchmark model. Consequently, the sub-optimal performance of ``impute, then predict" strategy-based models aligns with expectations, given their disregard for uncertainty surrounding missing values. It is worth noting that this strategy will lead to biased estimated parameters whatever imputation methods are used, which is proved by \citet{josse2019consistency}.
Moreover, the proposed model can be viewed as an end-to-end model that bypasses the imputation procedure, compared to the ``impute, then predict'' strategy, This idea aligns with the concept of recently proposed value-oriented forecasts for operational use by \citet{zhang2023value,zhang2023contextual}, i.e., optimizing the ultimate goal rather than the intermediate results.

Both the proposed model and the ``universal imputation" strategy-based models rely solely on observations for parameter estimation. However, it's crucial to note that the ``universal imputation" strategy hinges on the MAR/MCAR assumption. While the retraining strategy is capable of obtaining unbiased parameter estimates, we acknowledge that the proposed model is an approximation of the retraining strategy.
Consequently, in cases 1 and 2 where the data adhere to the MCAR assumption, the ``universal imputation" strategy-based model performs comparably well to the proposed model. However, it is essential to highlight that the computational efforts involved in training and forecasting differ, which will be elaborated in Section 7.3. Moreover, the applicability of the proposed model and the ``universal imputation" strategy-based models diverges. The predictive performance of the ``universal imputation" strategy-based models degrades when confronted with MNAR data.

\subsection{Discussion on Non-crossing Constraints}
\begin{figure}[!ht]
\centering
\subfigure[Reliability diagrams]{\includegraphics[width=0.48\textwidth]{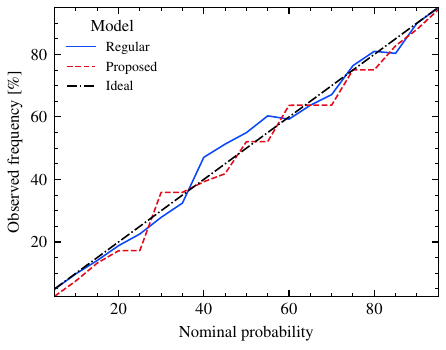}}
\subfigure[Sharpness diagrams]{\includegraphics[width=0.48\textwidth]{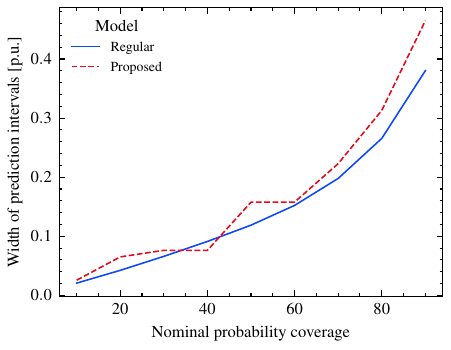}}
\caption{Assessment of 1-step ahead probabilistic forecasts for by the proposed and regular quantile regression model for Case 2, based on reliability diagrams (a) and sharpness diagrams (b).}
\label{fig_score_4}
\end{figure}

Unlike the approach of incorporating soft constraints on quantile crossing, as demonstrated by \citet{lu2022probabilistic}, we enforce strict monotonicity by placing hard constraints on quantiles. 
This is achieved by defining higher quantiles as the sum of lower quantiles and non-negative increments. The soft constraint method involves introducing a penalty hyper-parameter to account for the loss associated with quantile crossing. However, in practical scenarios, this may still result in quantile crossing phenomena, as tuning the penalty often requires manual efforts. In contrast, the proposed model inherently avoids embarrassing quantile crossing issues. Nevertheless, it's important to note that such constraints do impact parameter estimation. To elucidate, we depict reliability diagrams and prediction interval widths for both regular quantile regression and the proposed model based on the specifications outlined in case 2, as illustrated in Figure~\ref{fig_score_4}. The outcomes reveal that the regular quantile regression model exhibits a quantile-crossing phenomenon, accompanied by wider prediction interval widths compared to those of the proposed model. Furthermore, as the proposed model constructs quantiles through sequential addition operations, it is computationally restricted to sequential calculations, whereas the regular quantile regression model can simultaneously compute quantiles.

\subsection{Discussion on Model Complexity}
Instead of evaluating the algorithmic complexity of each method, we assess the training time for each model, a crucial consideration for forecasters. The training times for all models on CPU are presented in Table~\ref{table_4}, with the time spent on imputation for the ``impute, then predict" strategy-based models excluded. \revise{We note that the training of IM-Gaussian and DeepAR can be sped up by running on GPUs. In contrast, as discussed in Section 7.2, the hard constraints restrict the proposed model from parallel computation on GPUs; thus it is worth to investigate other non-crossing quantile regression neural network architectures that allow parallel computation.}
\revise{But the training efficiency of the proposed model is still affordable in practice, compared to the ``impute, then predict'' and ``universal imputation'' strategy based models}. It is evident that the proposed model exhibits efficiency in training, comparable to commonly used quantile regression models. Notably, the training time of the UI model (based on fully conditional specification) scales linearly with dimensionality, necessitating the training of a specific model for each conditional distribution. Additionally, the UI model relies on marginalization at the forecasting stage, which remains more computationally demanding than the proposed model.

\begin{table}[!ht]
\renewcommand{\arraystretch}{1.3}
\caption{The training time of the proposed and benchmark models in case 1 (minutes).}
\label{table_4}
\centering
\begin{threeparttable}
\begin{tabular}{cccccccc}
\hline
Models & Climatology & IM-Gaussian\tnote{*} & IM-QR & DeepAR\tnote{*} & UI & R-QR & Proposed\\
\hline
Training time & - & 32 & 10 & 64 & 41 & 10 & 6\\
\hline
\end{tabular}
\begin{tablenotes}
 \footnotesize
 \item[*] \revise{The training of neural network based models can be sped up via parallel computation.}
 \end{tablenotes}
\end{threeparttable}
\end{table}

\section{Conclusion}
Unlike the common practice of treating missing values in pre-processing through deletion and imputation, we propose a resilient probabilistic wind power forecasting approach within the commonly used conditional distribution modeling framework. Inspired by the fact that for each missingness pattern, we can estimate a specific conditional distribution, we design a forecasting model with parameters adaptive to missingness patterns. Then, it is only required to develop one model, but the model is applicable to all missingness patterns. Particularly, it is based on deep neural network models and includes a feature extraction module with biases adaptive to missingness patterns by design. Additionally, it has a non-crossing quantile neural network module responsible for yielding quantiles. 
Compared to the existing ``impute, then predict" strategy, the proposed approach is free of any pre-processing such as deletion and imputation, and therefore avoids introducing potential errors therein. In contrast to the ``universal imputation" strategy, the applicability of the proposed model is broader, i.e., it is applicable to cases under all missingness mechanisms. And it is computationally efficient, which is easy to implement in practice. Case studies demonstrate that the proposed approach achieves state-of-the-art performance in terms of CRPS at both missing-completely-at-random and missing-not-at-random cases and is time-efficient during training. 

However, we have only designed adaptive bias in feature extraction module; further endeavors could focus on designing adaptive weights. Additionally, this study assumes that parameters remain time-invariant, which may not align with real-world scenarios. Investigating situations where parameters exhibit time-variability would be valuable.

\appendix
\section{Implementation Details}
The hyperparameters for each model are listed as below.
\begin{description}
\item IM-Gaussian: The MissForest model for imputation is implemented via scikit-learn\footnote{https://scikit-learn.org/stable/} package, where the iteration is set as 100. The Gaussian distributional probabilistic forecasting model is implemented via a neural network in pytorch\footnote{https://pytorch.org/}. The neural network contains two hidden layer, each of which contains 512 units. The output layer yields the mean and variance of the Gaussian distribution.
\item IM-QR: The MissForest model for imputation is implemented via scikit-learn package, where the iteration is set as 100. The quantile regression model is implemented via gradient boosting regression trees within the scikit-learn package. The number of trees is set as 250, whereas the maximum depth is set as 5. The minimum samples per leaf is set as 9.
\item DeepAR: The DeepAR is implemented via GluonTS\footnote{https://ts.gluon.ai/stable/}. The model contains two recurrent neural network layers, each of which contains 40 units.
\item UI: The UI model is implemented via miceforest\footnote{https://pypi.org/project/miceforest/}. Specifically, the gradient boosting regression tress are used for predictive matching. The number of trees is set as 250, whereas the maximum depth is set as 5. The minimum samples per leaf is set as 9.
\item R-QR: The complete dataset is used. The quantile regression model is implemented via gradient boosting regression trees within the scikit-learn package. The number of trees is set as 250, whereas the maximum depth is set as 5. The minimum samples per leaf is set as 9.
\item The proposed model: It is implemented via a neural network in pytorch, where the feature extraction module contains 10 layer whereas the quantile regression module contains 3 layer. Each layer in the quantile regression module contains 512 units.
\end{description}

\section*{Acknowledgement}
This work is funded by National Natural Science Foundation of China (52307119). The author appreciates the constructive suggestions by anonymous reviewers.

\section*{Declaration of generative AI and AI-assisted technologies in the writing process}

During the preparation of this work the author used ChatGPT in order to improve language and readability. After using this tool, the author reviewed and edited the content as needed and takes full responsibility for the content of the publication.

\bibliographystyle{elsarticle-harv} 
 \bibliography{cas-refs,paper}





\end{document}